\documentclass[9pt,twocolumn,twoside]{optica}
\setboolean{shortarticle}{false}
\setboolean{minireview}{false}

\title{Anti-parity-time Symmetry in Passive Nanophotonics}

\author[1,2]{Heng Fan}
\author[1,2]{Jiayang Chen}
\author[1,2]{Zitong Zhao}
\author[3]{Jianming Wen}
\author[1,2,*]{Yu-Ping Huang}
\affil[1]{Physics Department, Stevens Institute of Technology, Castle Point at Hudson, Hoboken, NJ, 07648, USA}
\affil[2]{Center for Quantum Science and Engineering, Stevens Institute of Technology, Castle Point at Hudson, Hoboken, NJ, 07648, USA}
\affil[3]{Department of Physics, Kennesaw state University, Marietta, Georgia 30060, USA}

\affil[*]{Corresponding author: yuping.huang@stevens.edu}

\begin{abstract}
Parity-time (PT) symmetry in non-Hermitian optical systems promises distinct optical effects and applications not found in conservative optics. Its counterpart, anti-PT symmetry, subscribes another class of intriguing optical phenomena and implies complementary techniques for exotic light manipulation. Despite exciting progress, so far anti-PT symmetry has only been realized in bulky systems or with optical gain. Here, we report an on-chip realization of non-Hermitian optics with anti-PT symmetry, by using a fully-passive, nanophotonic platform consisting of three evanescently coupled waveguides. By depositing a metal film on the center waveguide to introduce strong loss, an anti-PT system is realized. Using microheaters to tune the waveguides' refractive indices, striking behaviors are observed such as equal power splitting, synchronized amplitude modulation, phase-controlled dissipation, and transition from anti-PT symmetry to its broken phase. Our results highlight exotic anti-Hermitian nanophotonics to be consolidated with conventional circuits on the same chip, whereby valuable chip devices can be created for quantum optics studies and scalable information processing.
\end{abstract}

\setboolean{displaycopyright}{true}

\begin{document}

\maketitle

\section{Introduction}

Optical realizations of non-Hermitian Hamiltonians possessing parity-time (PT) symmetry\cite{doi:10.1142/S0219887810004816,PhysRevLett.80.5243} promise innovative techniques such as enhanced sensing at exceptional points \cite{Hodaei2017, Lai_nature2019}, unidirectional light propagation \cite{PhysRevLett.106.213901,Feng2013}, and PT-symmetric laser \cite{Hodaei975,Feng972}. Exploiting the mathematical equivalence between quantum Schrodinger and electromagnetic paraxial equation, PT-symmetric optical systems have been constructed by balancing gain and dissipation \cite{EI2007,El-Ganainy2018,Feng2017,Wen_2018,RevModPhys.88.035002,ozdemir2019}. Its counterpart, anti-PT symmetry, represents another striking class of non-Hermitian physics \cite{Ying_science_2019,PhysRevLett.124.053901,PhysRevLett.123.193604}. Unlike PT-symmetry accompanied by real eigenvalues, anti-PT systems have purely imaginary eigenvalues, which give rise to surprising effects and extraordinary utilities \cite{Ying_science_2019}. Also, anti-PT can be realized without optical gain, which has been a source of noise and/or significant device overhead in PT systems. These, and other advantages, make anti-PT symmetry effects more accessible in practice and provide the prospect of mass integration with other optical components for scalable applications \cite{El-Ganainy2018,Feng2017,Wen_2018,RevModPhys.88.035002,ozdemir2019}. 
Thus far, effects of anti-PT symmetry have been observed in cold atoms \cite{PhysRevLett.123.193604,Peng2016}, electrical circuits \cite{Choi2018}, and very recently integrated photonics with balanced loss and gain induced by Brillouin scattering \cite{PhysRevLett.124.053901}. 

In this work, we report an all-passive realization of anti-PT optics on a versatile nanophotonic platform of thin-film lithium niobate on insulator (LNOI). Using three waveguides with carefully-engineered lateral coupling and loss, we observe the anti-PT symmetry and its transition to a broken phase via thermal-optical tuning. Incorporating it in an imbalanced Mach-Zehnder interferometer, we further achieve exotic synchronized amplitude modulation and phase-controlled dissipation. All of our experimental results are in good agreement with the numeric solutions of a non-Hermitian Hamiltonian, which affirms the anti-PT effects observed. Unlike previous realizations, here the system is constructed using purely passive optics without any optical gain \cite{PhysRevLett.123.193604,PhysRevLett.124.053901,PhysRevLett.103.093902}. Furthermore, the whole structure is monolithic etched using standard LNOI fabrication recipes, by which a variety of other optical elements, such as electro-optical modulators\cite{Cheng_2018_EOM, Mingwei_OL}, frequency converters\cite{Jia_2019_optica, Lu_2019_optica}, filters, and photon sources \cite{Ma2020}, can be integrated on the same chip. This highlights the prospect of creating exotic chips that consolidate coherent, non-Hermitian, and anti-Hermitian optical circuits, for applications in areas of photonic computing, communications, sensing, and so on. 
%owing to its high flexibility on fabrication and spatial arrangement of gain and loss elements. PT symmetry has been manifested in many systems in nano-photonics,including waveguide\cite{PhysRevLett.103.093902}\cite{Ruter_Natphys}, microring resonator\cite{Peng_natphys_2014}\cite{PhysRevLett.116.203902}, and photonic crystal cavity\cite{zhen_nature2015}\cite{PhysRevLett.116.203902}\cite{PhysRevB.92.235310}. Exotic phenomenons,which might be prohibited by hermitian physics when comes to optical loss, presented in those systems have paved new paths to manipulate optics. Lithium niobate thin film on insulator(LNOI), as a emerging powerful platform in nano-photonics\cite{Mian_2017_optica}, has demonstrated the highest single photon non-linearity on chip recently\cite{Jia_2019_optica}\cite{Lu:19}. Beyond that, LNOI has moderate optical performance overall, including  electro-optical modulation\cite{Cheng_2018_EOM}\cite{Mingwei_OL} and piezoelectricity\cite{Linbo_2019_optica}, which turns LNOI to be a revolutionizing candidate for both optical communication and computation\cite{Andreas_2018}. Still, LNOI based photonic integrate circuits are short of rich steering tools, compared to electrical circuits. One of main goals of non-Herminian physics is to deliver novel manipulation methods on optics. We expect optical circuit integrates non-hermitian elements,implemented in our paper, with aforementioned active components pave integrate photonic circuits to a upper stage.

\section{Structure design and Model}
\label{sec:examples}

\begin{figure}[htbp]
\centering
\fbox{\includegraphics[width=\linewidth]{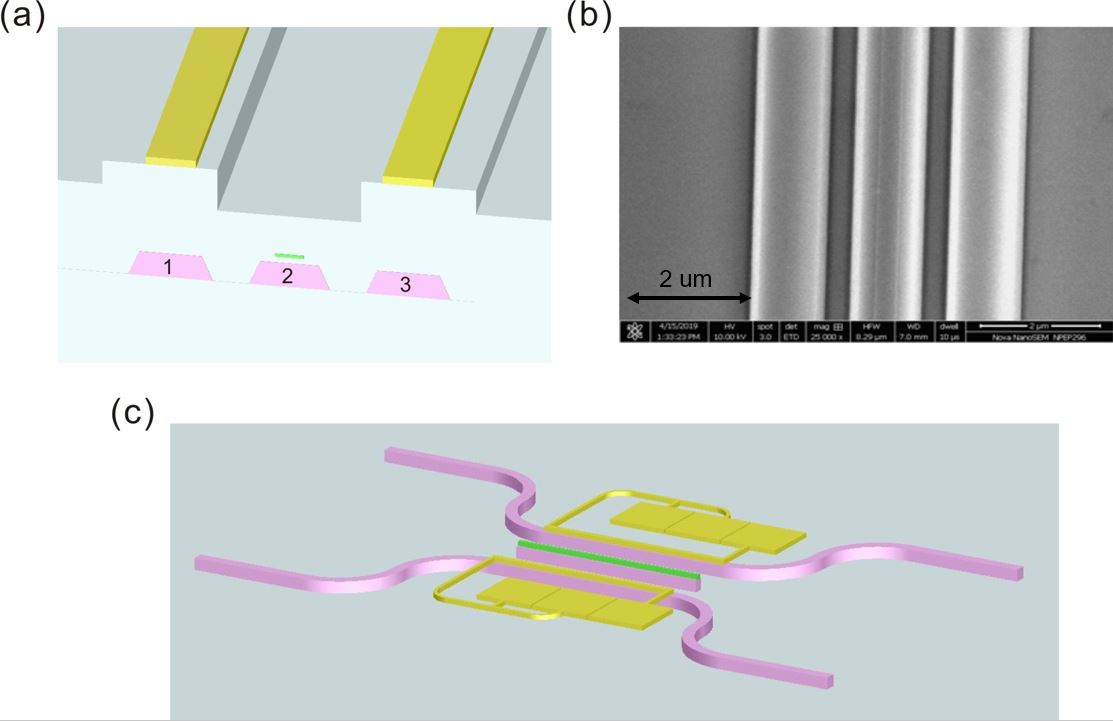}}
\caption{Anti-PT structure with three laterally coupled waveguides. (a) and (b) show the side and top views, and (c) illustrates the layout. The top gap between the Waveguides is around 820 nm. The width and thickness of the middle trench are 1.9 $\mu$m and 690 nm, respectively. Its side wall, etched through ICP-FL, is at an angle close to 90 degree.}
\label{Images: Fig_1}
\end{figure}

A schematic of the anti-PT nanophotonic circuit is depicted in Fig.~\ref{Images: Fig_1}, which implements the proposal in Ref. \cite{PhysRevA.96.053845,Wen_2018}. It consists of three evanescently coupled waveguides, labeled as $1$, $2$, and $3$. They are fully etched by ion milling on a Z-cut lithium niobate thin film bonded on silicon oxide above a silicon substrate. %Measured under a scanning electron microscope (SEM),
%and 8 870 nm for Waveguide 1 and 3, and 820 nm for Waveguide 2, while the height are 400 nm for all.
The top widths of three waveguides are measured to be 870, 820 and 870 nm, respectively, and their heights are all 400 nm. Waveguide $2$ is cladded with a chromium strip to introduce strong propagation loss, which is measured at an amplitude decay rate of 1080 dB/cm. The details of device nanofabrication and characterization are presented in the Supplementary Material.

The system's wave dynamics is described by following coupled mode equations under the solwly varying envelope approximation \cite{PhysRevA.96.053845}:
\begin{equation}
\begin{split}
\frac{da_1}{dz} &=-i(\Delta/2)\, a_1-i\kappa\, a_2
\\
\frac{da_2}{dz} &=-i\Delta'\,a_2-i\kappa^{*}(a_1+a_3)-\gamma \, a_2
\\
\frac{da_3}{dz} &=i(\Delta/2)\,a_3-i\kappa\, a_2
\end{split}
\end{equation}
where $a_{m}$ ($m=1,2,3$) is the mode amplitude in Waveguide $m$. $\kappa$ is the lateral coupling rate between neighboring waveguides. $\gamma$ is the dissipation rate of the Waveguide 2 mode induced by the chromium strip. $\Delta=k_1-k_3$ and $\Delta'=(k_1+k_3)/2-k_2$ are each the relative propagation constants, thus the effective detuning, of the waveguide modes. 

To realize an effective anti-PT system, the dissipation in Waveguide 2 must be predominantly large, i.e., $\gamma \gg |\Delta'|,\,|\kappa|$. Under this condition, adiabatic elimination can be applied to reduce the equations of motion as
\begin{align}
i\frac{\partial}{\partial z}\begin{bmatrix}
a_1\\a_3
\end{bmatrix}= H_\texttt{eff}\begin{bmatrix}
a_1\\ a_3
\end{bmatrix}, 
\end{align}
where the effective Hamiltonian
\begin{equation}
\label{eq3}
H_\texttt{eff}=\begin{bmatrix}
\Delta/2-i\Gamma & -i\Gamma \\ -i\Gamma
 & -\Delta/2-i\Gamma
\end{bmatrix},
\end{equation}
with $\Gamma=|\kappa|^2/\gamma$. It is clear that $H_\texttt{eff}$ is anti-Hamitian that anti-commutes with the parity-time operators $\hat{P}$ and $\hat{T}$, i.e., $\{\hat{P}\hat{T},H_\texttt{eff}\}=0$. Here $\Gamma$ is the effective ``imaginary'' coupling strength between Waveguide 1 and 3, which is distinct from typical PT systems where the coupling constants are real \cite{Ruter_Natphys, Peng_natphys_2014,Chang2014}. 

Unlike Hermitian Hamiltonians, $H_\texttt{eff}$ gives non-trivial eigen solutions. When $\Gamma>\Delta/2$, there are two eigenvalues 
\begin{equation}
   \lambda^{s}_\pm =-i(\Gamma \pm \sqrt{\Gamma^2-\Delta^2/4}),
   \label{eqn_pt}
\end{equation}
which are purely imaginary. The corresponding eigenstates are
\begin{equation}
    \Psi^{s}_{\pm}=\left(\pm e^{\pm i\phi},1\right)/\sqrt{2},
\end{equation}
with $\phi=\sin^{-1}(\Delta/2\Gamma$). Thus, both eigenstates correspond to the two waveguides sharing the equal amplitude, but with different relative phase. They are orthogonal with each other only when $\Delta=0$, in which case $\Psi_+$ experiences propagation decay at rate $2\Gamma$, but $\Psi_-$ is lossless. Hence, only $\Psi_-$ could survive after sufficient propagation length. The same implies synchronized output and equal splitting of input power that are robust against the structure's parameter errors such as the waveguide length, loss, and coupling. When $\Delta\neq 0$, the two eigenstates are not orthogonal with each other, with unequal propagation losses differed by $2\sqrt{\Gamma^2-\Delta^2/4}$.

On the other hand, when $\Gamma<\Delta/2$, the eignvalues are no longer purely imaginary, becoming 
\begin{equation}
    \lambda^b_\pm =-i\Gamma \pm \sqrt{\Delta^2/4-\Gamma^2},
    \label{eqn_ptbroken}
\end{equation}
The system is then in a broken phase of the PT anti-symmetry, whose eigenstates are 
\begin{equation}
    \Psi^b_{\pm}=(ie^{\pm r},1),
\end{equation}
with $r=\cosh^{-1}(\Delta/2\Gamma)$. Unlike the previous case, now the distribution of optical power is asymmetrical over the two waveguides. Under this circumstance, this system exhibits non-reciprocal behaviors, and the optical power will mostly remain in the incident waveguide when $\Delta$ is large \cite{PhysRevA.96.053845}.  

On our chips, two microheaters are deposited 1.5~$\mu$m above Waveguide 1 and 3, respectively, to precisely tune $\Delta$ via thermo-optic effects while inducing only negligible loss. To utilize the highest thermo-optic coefficient of LNOI, fundamental transverse-magnetic (TM) modes are employed for all waveguides. To further increase the tunability, a trench is also etched above the center waveguide to partially block thermal flow. These allow us to flexibly tune $\Delta$ without affecting other system parameters, including the waveguide mode profiles and coupling strength. With a maximum of 200 mW electric power applied to either microheater, we are able to sweep $\Delta$ and observe the phase transition from the symmetry to broken phase, and vice versa; see Supplementary Material.

\section{Results and discussion}
In our setup, a polarized beam at 1550~nm is coupled into a fundamental TM mode of Waveguide 3 though an objective lens (C330-TMD-C from Thorlabs) as the input. The output light, from Waveguide 1 or 3, is collected by a lensed fiber (OZ optics) placed on a translation stage, and subsequently measured using a power meter. A varied electric current is applied to either of the two microheaters to tune $\Delta$. 

\begin{figure}[htbp]
\centering
\fbox{\includegraphics[width=\linewidth]{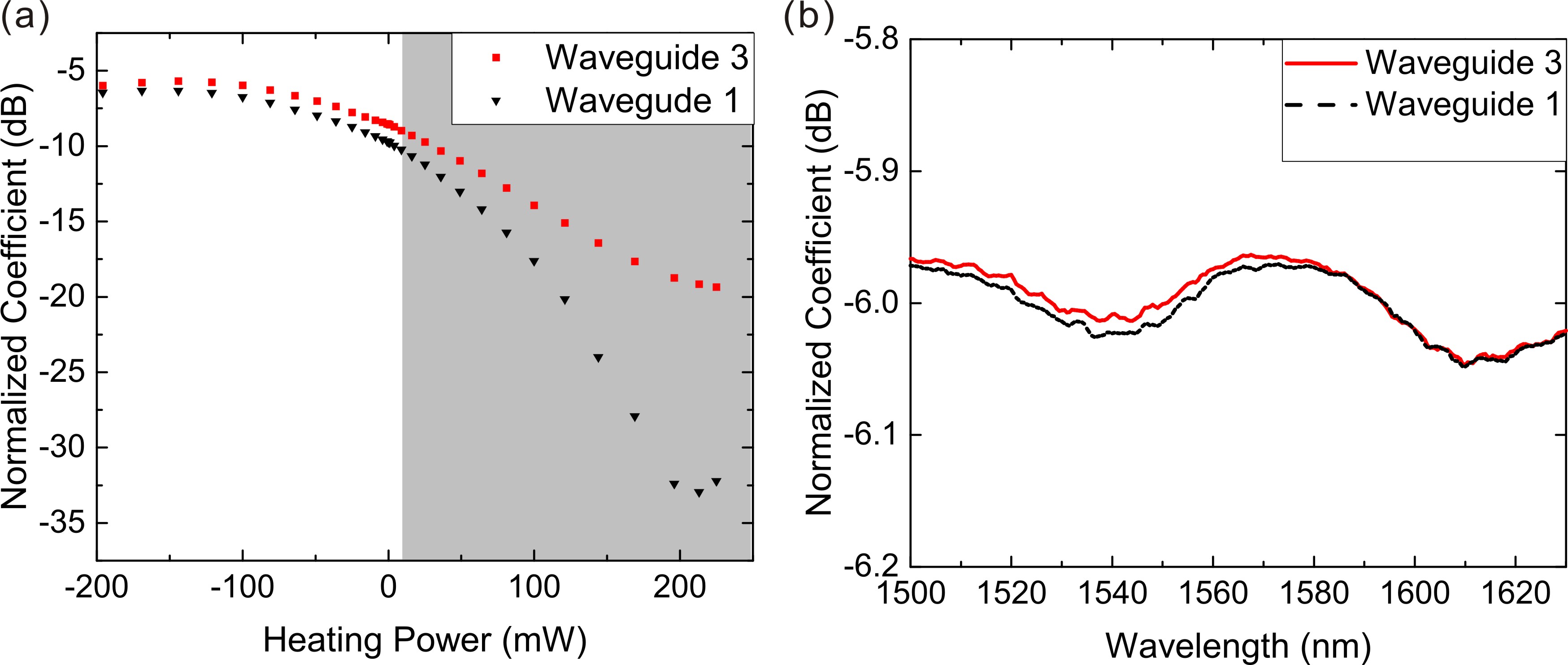}}
\caption{(a) Waveguide output power as a function of the applied heating power. Negative and positive heating power corresponds to the electrical current applied to the microheaters above Waveguide 1 and 3, respectively. Grey region marks the symmetry broken phase. (b) Power equal splitting over scanning wavelengths. Both figures have been normalized by the total coupling efficiency, with the input power scaled to 1 mW.}
\label{Images: Fig_2}
\end{figure}

In the first experiment, increasing current is applied to the microheater above Waveguide 1, which leads to increased optical power at the output of both waveguides, signifying a decreased detuning. As the applied heating power reaches 135~mW, the output optical power from the two waveguides is maximized simultaneously. At this point, $\Delta$ is close to zero. Afterwards, the current is switched to the microheater on Waveguide 3. The output power of both waveguides plummets as expected, due to an increased $\Delta$. The output power from Waveguide 1 is minimized when the heating power reaches 201~mW. The experimental results are summarized in Fig.~\ref{Images: Fig_2}(a), where each waveguide' output power is plotted as a function of the applied heating power. In the figure, the negative and positive heating power correspond to the current applied to the microheaters on Waveguide 1 and 3, respectively. From these data, the overall coupling efficiency, attributed to the losses of the fiber coupler, the objective lens to the waveguide, and the waveguides to the lensed fiber, is extract to be -14.8~dB by fitting the data in Fig.~\ref{Images: Fig_2}. The inter-waveguide coupling strength $\kappa$ and detuning $\Delta$ can also be extracted, while the induced dissipation rate $\gamma$ is measured using a cut-back method; see Fig. S3 and Table. S1 in the Supplementary Material for details. 

As shown in Fig.~\ref{Images: Fig_2}, the output power from the two waveguides is almost equal when the effective detuning approaches zero. It thus invites to construct a 50:50 beamsplitter on chip using the present anti-PT structure. Distinct to directional coupling \cite{directionalcoupler}, such equal beam splitting realized under anti-PT symmetry is not sensitive to the inter-waveguide coupling strength, the waveguide length, its propagation loss, or the optical wavelength. To demonstrate this, Waveguide 1 is applied with 135~mW heating power to bring the effective detuning close to zero. Then, the output power from Waveguide 1 and 3 are measured as the input laser wavelength is swept over a spectral range of 130~nm, as allowed by our laser tuning range. The results are shown in Fig.~\ref{Images: Fig_2}(b), where very flat, equal power splitting is observed over the entire spectrum, with less than $\pm 0.05 dB$ deviation. As such equal splitting only requires $\Gamma \gg \Delta, 1/L$ ($L$ is the waveguide length), it greatly boosts the tolerance of nanofabrication errors as compared with a coherent-optical circuit; see Fig. S5 in the Supplementary Material. Furthermore, this operational bandwidth is much wider than typically achievable with beam-splitters based on multimode interference or directional coupling \cite{MMI2011}. It implies practical values for applications with ultra-broadband signals and could prove useful for mass chi-integration.

\begin{figure}[htbp]
\centering
\fbox{\includegraphics[width=\linewidth]{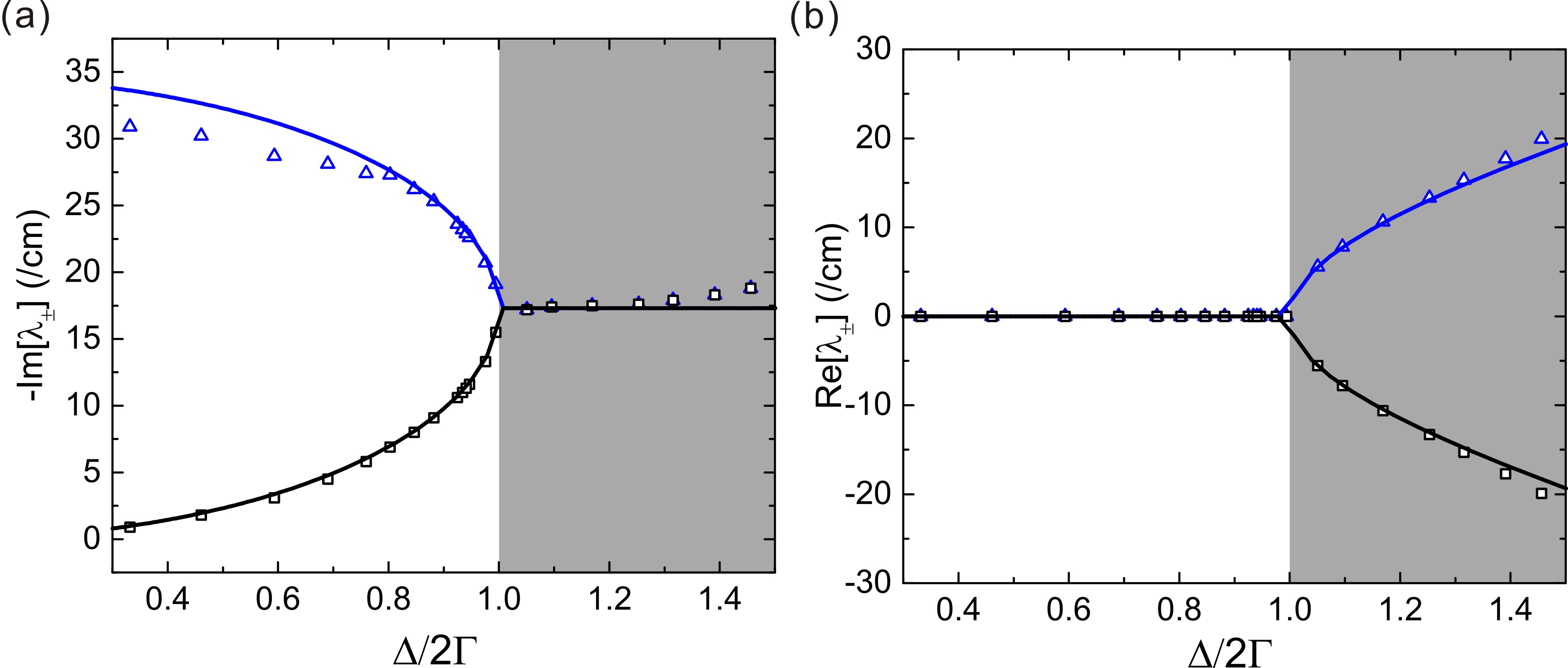}}
\caption{(a) and (b): Imaginary and real parts of the eigenvalues. Squares and triangles are from the experimental data, while the black and blue lines are theoretical results. Grey regions mark the symmetry broken phase.}
\label{Images: Fig_3}
\end{figure}

Using the coupling efficiency from the previous fitting results, the coupling strength and the effective detuning can be evaluated for each heating power by Eqs.~(S3) and (S4) in Supplementary Material. As expected, $\kappa$ is nearly independent of the applied heating power, while $\Delta$ increases linearly with it, as shown in Fig.~S3 and S4 of Supplementary Material. Then the eigenstates and eigenvalues of the effective Hamiltonian (\ref{eq3}) can be calculated at each heating power point. The results are given in Fig.~\ref{Images: Fig_3}, where the imaginary and real parts of the eigenvalues are plotted as functions of $\Delta/2\Gamma$. On the same figure, the theoretical results using $\kappa=136$~cm$^{-1}$, obtained by averaging over the various heating power applied, and $d\Delta/dp=0.22$~cm$^{-1}$mW$^{-1}$, as determined by the fitting in Fig.~S4 of Supplementary Material. As shown, there is a clear phase transition occurring at the exceptional point of $\Gamma=\Delta/2$, as predicted by Eqs.(\ref{eqn_pt}) and (\ref{eqn_ptbroken}). When $\Gamma>\Delta/2$, the two eigenvalues are purely imaginary, whose amplitude difference increases to nearly 30/cm as the detuning is decreased to $\Delta=0.6 \Gamma$. Meanwhile, the real wave vectors for the two eignmodes remain the same. When $\Gamma <\Delta/2$, in contrast, the two eigenstates share the same dissipation rate but have different wave vectors. 

\begin{figure*}[t]
\centering
\fbox{\includegraphics[width=\linewidth]{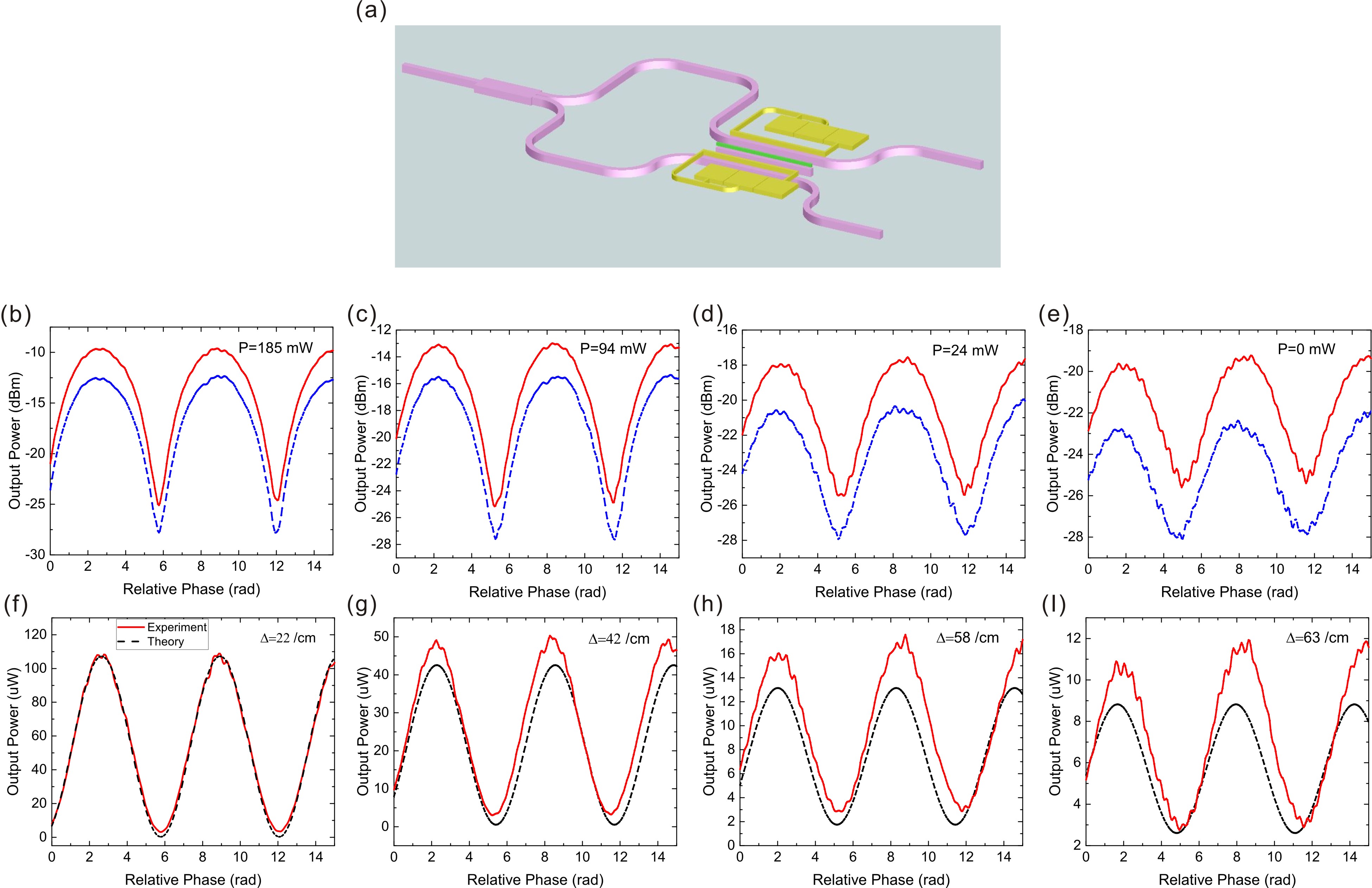}}
\caption{(a) Anti-PT structure in an imbalanced Mach-Zehnder interferometer. (b)-(e) Output power from two waveguides as the input relative phase, $\theta$, varies, with different heating power $P$ applied to Waveguide 1. Red curves are for Waveguide $3$, and Blue dash lines are for Waveguide $1$. (f)-(i) Comparison of the measured Waveguide 3 output with theory. $\theta$ is calibrated by observing its linear dependence of the laser wavelength and the 2$\pi$ period of a full oscillation}
\label{Images: Fig_4}
\end{figure*}

Another distinct feature is the output power synchronization: with two inputs of equal amplitude, the output power of the two waveguides is always equal, regardless of the relative phase between the two inputs. This holds irrespective of whether the system is in the anti-PT symmetry or its broken phase. On the other hand, the total output power is not conserved and sensitive to the relative phase. Such a phenomenon is in direct contrast to that in coherent-optical couplers describable by Hermitian Hamiltonians, where the total output power is conserved but its distribution over the outputs is sensitive to the input relative phase. This effect thus complements what coherent optics can offer, which may find utilities applications in areas of sensing\cite{Lai_nature2019}, measurement, optical modulation, and lasers. 
To observe those effects, we fabricate a LNOI circuit incorporating the anti-PT structure into an imbalanced interferometer, as shown in Fig. \ref{Images: Fig_4}(a). An input laser is first split equally into two parts upon a multimode interferometer. They are then each guided along two waveguides of the same cross section but different path length, before fed into an input. Their relative phase acquired before the anti-PT structure is given by $\theta=n d_L/\lambda$, with $n$ being the waveguides' refractive index, $d_L$ the path length difference, and $\lambda$ the laser wavelength. In our structure, $d_L=435$ $\mu$m, so that by scanning $\lambda$ from 1544 nm to 1550 nm, $\theta$ is swept over 16 radians nearly linearly with $\lambda$. Meanwhile, other system parameters, such as $\Gamma$ and $\Delta$, remain practically unchanged to allow observing the effects caused only by the relative phase variation.  

The results are presented in Figs. \ref{Images: Fig_4}(b-e), where the output power from the two waveguides is measured individually as $\lambda$, thus $\theta$, is swept under diffphase between subsequent peaks is $2\pi$. As seen, in all cases, the two outputs are always equal in power except for a constant factor ($\sim$ 2 dB), which is likely attributed to the difference in the waveguide propagation and coupling loss in the two outputs, caused by, e.g., stitching errors during nanofabrication. The total output power, on the other hand, oscillates with $\theta$, where a high 15~dB extinction is achieved with 185~mW heating power applied to the microheater above Waveguide 1, as seen in Fig.~\ref{Images: Fig_4}(b). These results demonstrate clear effects of synchronized amplitude modulation and phase-controlled dissipation.  

To understand the above phenomena, we simulate the system dynamics using Hamiltonian (\ref{eq3}), with $\Gamma=17$ cm$^{-1}$, the coupling efficiency, and other parameters extracted from the previous measurements in Fig.2. As a comparison, Figs. \ref{Images: Fig_4}(f-i) plot the measured and simulated output power from the Waveguide 3. In all four figures, the only fitting parameter used is the baseline detuning $\Delta$, as it is extremely sensitive to the waveguide's cross section and cannot be otherwise determined. Rather, $\Delta$ is best fitted to be $22$ cm$^{-1}$ when the heating power is 185 mW, as shown in \ref{Images: Fig_4}(f). For other heating settings, $\Delta$ is calculated directly using its linear dependence of the applied heating power. As shown, in all cases, the experimental and simulation results agree fairly well. In Fig.~\ref{Images: Fig_4}(f), $\Delta/2\Gamma=0.65$, so that the system is in anti-PT symmetry. The output power thus experiences the strongest loss when the two inputs are in phase and the least loss when they are out of phase. This is expected because in the $\Gamma \gg \Delta/2$ limit, each of the two cases correspond to $\Psi^s_+$ and $\Psi^s_-$, whose eigenvalues differ by $-2i\sqrt{\Gamma^2-\Delta^2/4}$. Thus their losses through the structure are significantly different, giving rise to the high extinction as $\theta$ is varied. As $\Delta$ increases, the propagation loss difference in the two cases decreases, so is the extinction. In Fig.~\ref{Images: Fig_4}(i), when $\Delta/2\Gamma =1.85$, the system is significantly into the broken phase, for which the extinction drops to 5 dB. 

\section{Conclusion}

In conclusion, We have observed anti-PT symmetry and its transition into a broken phase using an all-passive, nanophotonic platform. By thermal tuning, we observed striking phenomena such as power equal splitting, synchronized amplitude modulation, and phase-controlled dissipation under anti-PT symmetry, as well as their behaviors as the system evolves passing the exceptional point and into the broken phase. The good agreement between our experimental results with the predictions of a non-Hermitian Hamiltonian affirms the anti-PT effects observed, which are not found in other conservative or PT-symmetric optics. They could be used to construct synchronized modulators, novel optical interferometers, and nontrivial logical gates, such as exclusive-OR phase gate. By replacing the thermal tuning with electronic tuning to control the relative phase, one might also realize exotic electro-optical interfaces as well. 

As the anti-PT structure is realized entirely using purely passive optics and monolithic etched on lithium niobate thin films, hybrid chips are ready to be developed incorporating Hermitian and anti-Hermitian circuits for exotic functionalities. Our work complements the exciting progress made recently in lithium niobate nanophotonics, as such electro-optical modulators, frequency converters, filters, and photon sources. By consolidating a variety of other elements and circuits on the same chip, exceptional optical devices and systems can be developed for photonic computing, communications, sensing, and so on.

\section*{Funding Information}
This research is supported in part by National Science Foundation (Awards: 1806523, 1806519, and EFMA-1741693).

%\section*{Acknowledgments}

\section*{Disclosures}

\medskip

\noindent\textbf{Disclosures.} The authors declare no conflicts of interest.

\section*{Supplemental Documents}
See Supplement 1 for supporting content.

%\bigskip \noindent See \href{link}{Supplement 1} for supporting content.

% Bibliography
\bibliography{sample}

% Full bibliography will be added automatically on a new page for Optics Letters submissions. This command is ignored for journal article submissions.
% Note that this extra page will not count against page length.
\bibliographyfullrefs{sample}

%Manual citation list
%\begin{thebibliography}{1}
%\bibitem{Zhang:14}
%Y.~Zhang, S.~Qiao, L.~Sun, Q.~W. Shi, W.~Huang, %L.~Li, and Z.~Yang,
 % \enquote{Photoinduced active terahertz metamaterials with nanostructured
  %vanadium dioxide film deposited by sol-gel method,} Opt. Express \textbf{22},
  %11070--11078 (2014).
%\end{thebibliography}

\end{document}

\maketitle

\section{Device fabrication}

The devices are fabricated on a 400-nm thick Z-cut thin film lithium niobate on insulator (LNOI), based on 1.8-$\mu$m silicon oxide (NANOLN Inc.). The wafer is spin-coated with HSQ resist (Fox 16) and baked at 90 degree for 4 minutes. The first electron beam lithography step is conducted to pattern waveguides and alignment marks. Two-pass writing is implemented to increase the uniformity especially in the waveguide coupled region. The wafer is developed by TMAH (25$\%$) and etched through ion-milling process \cite{Chen:18,Chen:19}. Those waveguides are fully etched, then cleaned with RCA-I solution at 60 degree. The top width and base width of waveguides are measured under a scanning electron microscope. A layer of 100-nm silicon oxide is deposited on the wafer using PECVD. The wafer is cleaned and spun with one layer of PMMA 495 A6 and PMMA 950 A4, respectively. The second lithography step is conducted to define strips on top of the central waveguides of those three coupled waveguides structures. A layer of 20-nm chromium is deposited on the wafer using electron-beam evaporator after development, followed by a lift-off process in Remover PG at 80 degree. The wafer is cleaned and cladded with 1.4 um silicon oxide. The third lithography is conducted to define trenches, for thermal isolation, over the central waveguides with a layer of PMMA 950 A11 film. The wafer is then developed by IPA plus DI water and etched by ICP-FL. The depth of trench is 690-nm, measured on a referred trench on the same wafer though stylus. The final lithography step is conducted to define microheaters with two layers of PMMA 495 A6 and one layer of PMMA 950 A4. The wafer is developed by DI water plus MIBK and deposited with 30-nm of Cr and 100-nm of gold though electron-beam evaporator. An overnight lift-off process is implemented using Remover PG at 80 degree. Finally the wafer is diced and polished for measurement.

\section{Dynamics of three coupled waveguides structure}

The evolution of three coupled waveguides is governed by:
\begin{equation}
\label{S1}
\begin{split}
\frac{da_1}{dz}&=-ik_1\, a_1-i\kappa \,a_2 
\\
\frac{da_2}{dz}&=-ik_2\,a_2-i\kappa^{*}(a_1+a_3)-\gamma\, a_2
\\
\frac{da_3}{dz}&=-ik_3\, a_3-i\kappa\, a_2
\end{split}
\end{equation}
where $k_m(m=1,2,3)$ are the propagation constants of modes, and $\kappa$ is the modal coupling strength. Owing to the identical dimensions of Waveguides $1$ and $3$ as designed, their lateral coupling strength with Waveguide 2 is treated as the same. $\gamma$ is the mode dissipation rate of the waveguide 2 induced by a chromium strip. In this paper, we only consider the fundamental quasi-transverse-magnetic (quasi-TM) modes in all waveguides and a wavelength at 1550 nm. When $\gamma$ is large, adiabatic elimination could be employed to reduce Eq.~\ref{S1} to dissipative coupling between $a_1$ and $a_3$ at strength $\Gamma=\frac{|\kappa|^2}{\gamma}$. When  $\gamma \gg |\Delta'|,\,|\kappa|$, an anti-PT Hamiltonian is realized effectively as 
\begin{equation}
i\frac{\partial}{\partial z}\begin{bmatrix}
a_1\\a_3 
\end{bmatrix}= \begin{bmatrix}
\Delta/2-i\Gamma & -i\Gamma \\ -i\Gamma
 & -\Delta/2-i\Gamma
\end{bmatrix}\begin{bmatrix}
a_1\\a_3 
\end{bmatrix}
\end{equation}
where $\Delta=k_1-k_3$ and $\Gamma$ are the effective detuning and coupling strength, respectively. The eigenvalues could be solved following Ref.~\cite{PhysRevA.96.053845} separately in two regions, with the phase transition happening at $\Gamma=\Delta/2$. With input power $P_\textrm{in}$ in Waveguide 3, the output power under anti-PT symmetry reads
\begin{equation}
\label{S3}
\begin{split}
P_1&=P_\textrm{in}e^{-2\Gamma L} \frac{\Gamma^2}{S^2}\sinh^2(SL)\eta_1,
\\
P_3&=P_\textrm{in} e^{-2\Gamma L}  (\cosh^2(SL)+\frac{\Delta^2}{4S^2}\sinh^2(SL))\eta_2,
\end{split}
\end{equation}
where $P_1$ and $P_3$ are the output power from Waveguide $1$ and $3$, respectively, with $\eta_1$ and $\eta_2$ each's total coupling efficiencies. $L$=0.1 cm is the length of the coupled region in our structure. $S$ is defined as $S=\sqrt{\Gamma^2-\frac{\Delta^2}{4}}$. 

In the symmetry broken phase where $\Gamma<\Delta/2$, the output power reads: 
\begin{eqnarray}
\label{S4}
\begin{split}
P_1&=P_\textrm{in}e^{-2\Gamma L} \frac{\Gamma^2}{q^2}\sin^2(qL)\eta_1,
\\
P_3&=P_\textrm{in}e^{-2\Gamma L} (\cos^2(qL)+\frac{\Delta^2}{4q^2}\sin^2(qL))\eta_2,
\end{split}
\end{eqnarray}
where $q=\sqrt{\frac{\Delta^2}{4}-\Gamma^2}$. The input power is set to be 1 $mW$. From those equations, $\Gamma$ and $\Delta$ can be solved independently under each heating power with known $P_1$, $P_3$, $L$, and $\eta_{1,2}$. Furthermore, the inter-waveguide coupling strength $\kappa$ can be obtained, as the induced loss $\gamma$ is determined independently by a cut-back method. Later in this Supplementary Material, we will fit experimental results to extract $\eta_{1,2}$. 

Next, we consider two inputs with the same amplitude and a relative phase ($\theta$) (assuming $\eta_1=\eta_2=\eta$)
\begin{eqnarray}
\begin{split}
 P_1=P_3=& P_\textrm{in} e^{-2\Gamma L}[1+\frac{2\Gamma ^2}{S^2}\sinh^2(SL)+\frac{\Delta\Gamma}{S^2}\sinh^2(SL)\sin(\theta)\\
&-\frac{\Gamma}{S}\sinh(2SL) \cos(\theta)]\eta \quad \text{ if } \Gamma >\Delta/2,    \\
P_1=P_3=& P_\textrm{in} e^{-2\Gamma L}[1+\frac{2\Gamma ^2}{q^2}\sin^2(qL)+\frac{\Delta\Gamma}{q^2}\sin^2(qL)\sin(\theta)\\
&-\frac{\Gamma}{q}\sin(2qL)\cos(\theta)]\eta  \quad \text{ if } \Gamma \leqslant \Delta/2. \\
\end{split}
\end{eqnarray}

\section*{Device characterization and results}

\begin{figure}[htbp]
\centering
\fbox{\includegraphics[width=\linewidth]{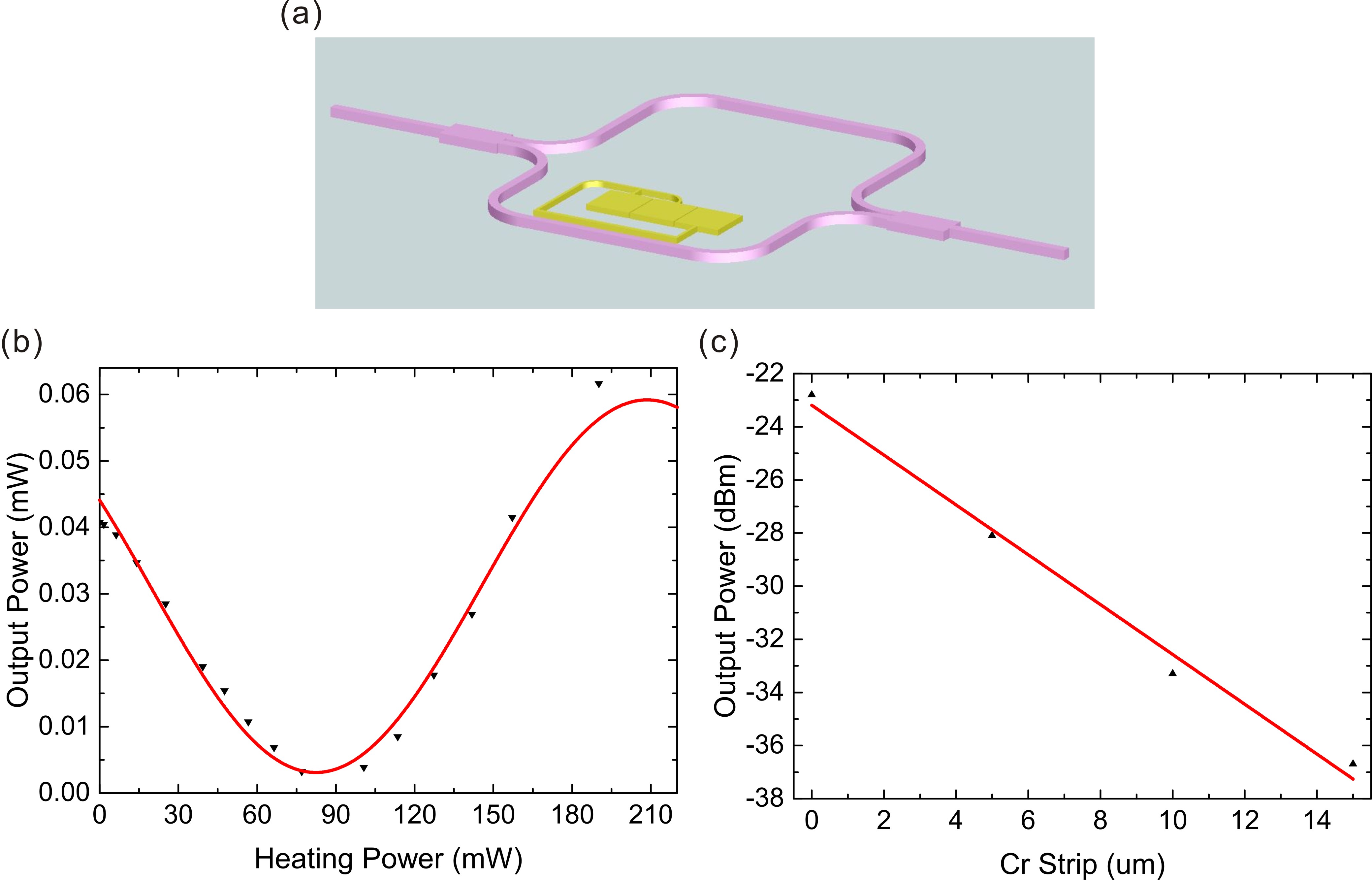}}
\caption{Microheater and amplitude decay rate calibration. (a) Layout of the Mach-Zehnder interferometer to measure the tunability of the microheater. The resistance is 630 $\Omega$, measured by a multi-meter. The length of the microheater is 1 mm. (b) Output power of the interferometer dependence of the heating power. (c) Amplitude decay rate measurement, black points and red line are experimental and fitted results, respectively.}
\label{fig:S1}
\end{figure}

In this experiment, we control the electric power applied to one of the  microheaters to tune $\Delta$ via thermo-optic effect. To characterize the relationship between the heating power and the propogation constants of the fundamental TM modes, a Mach-Zehnder interferometer with a microheater in one arm is employed, whose dimension is the same as in the anti-PT structure. As shown in Fig.~\ref{fig:S1}(a), the heating current is loaded onto two touch pads through two radio-frequency microwave probes (by Signatone). The shift in the wave propagation constant $k$ caused by the thermo-optic effect is linearly proportional to the heating power $P$. The output power of the interferometer is recorded as $P$ is varied gradually from 0 mW to 192 mW. Then, the thermo-optic coefficient is fitted as
\begin{equation}
\frac{d k}{d P}=\frac{2\pi\cdot\Delta n}{\lambda \cdot\Delta P} =24.9*10^{-2}/(mW\cdot cm),
\label{S6}
\end{equation}
as shown in Fig. \ref{fig:S1}(b). %Note that the phase transition is realized though sweeping the effective detuning, not the propagation constant of mode in a waveguide. Later, we will build the relationship between the heating power and the effective detuning.

To characterize the absorption loss strength on the fundamental TM mode induced by a chromium strip above the waveguides, we use the cut-back method and measure the loss for different strip length. The result is shown in Fig.~\ref{fig:S1}(c), where the amplitude decay rate is fitted to be 1080/cm. It agrees with the numeric simulation result of  $\gamma_{sim}=1184/cm$ using the Lumerical MODE solver.

\begin{figure}[htbp]
\centering
\fbox{\includegraphics[width=\linewidth]{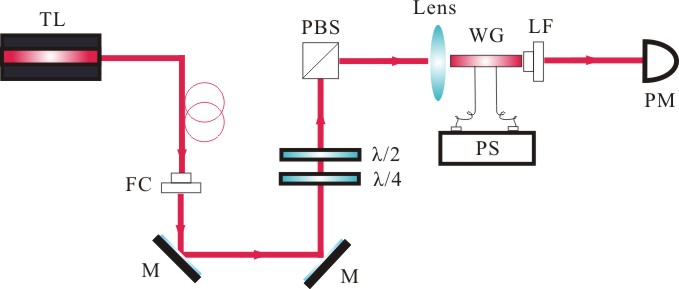}}
\caption{ Schematic of the experimental setup. TL: Tunable laser; FC: Fiber coupler; M: Mirror; $\lambda$/4: Quarter-wave plate; $\lambda$/2: Half-wave plate; PBS: Polarization Beam Splitter; PS: Power source; WG: Waveguide; LF: Lensed fiber; PM: Power meter.}
\label{fig:S2}
\end{figure}
The schematic of the experimental setup is depicted in Fig. \ref{fig:S2}.
A tunable telecom-band laser is aligned through a fiber collimator using molded glass aspheric Lenses (C220-TMD-C by Thorlabs). The reflection of a polarization beam splitter (PBS 201 by Thorlabs) is focused by the same lenses as in the fiber coupler, and injected into the waveguide. The power of the input laser is adjusted by a pair of half and quarter waveplates. The waveguide output is collected by a lensed fiber (spot diameter: 2.0$\pm$0.5 $\mu$m, by OZ optics), and then measured by a power meter. The reflected light from the PBS is injected into Waveguide $3$ which excites its quasi-TM mode. The input wavelength is 1550 nm and the input power is 10 dBm. Then, the output power from Waveguides $1$ and $3$ are measured as the heating power is varied. The loss of the fiber coupler is -4.6 dB. The total coupling loss of reference waveguides on the same wafer, including the fiber coupler loss and the waveguide input and output losses, is typcially measured to be around -14 dB.

\begin{figure}[ht]
\centering
\includegraphics[width=\linewidth]{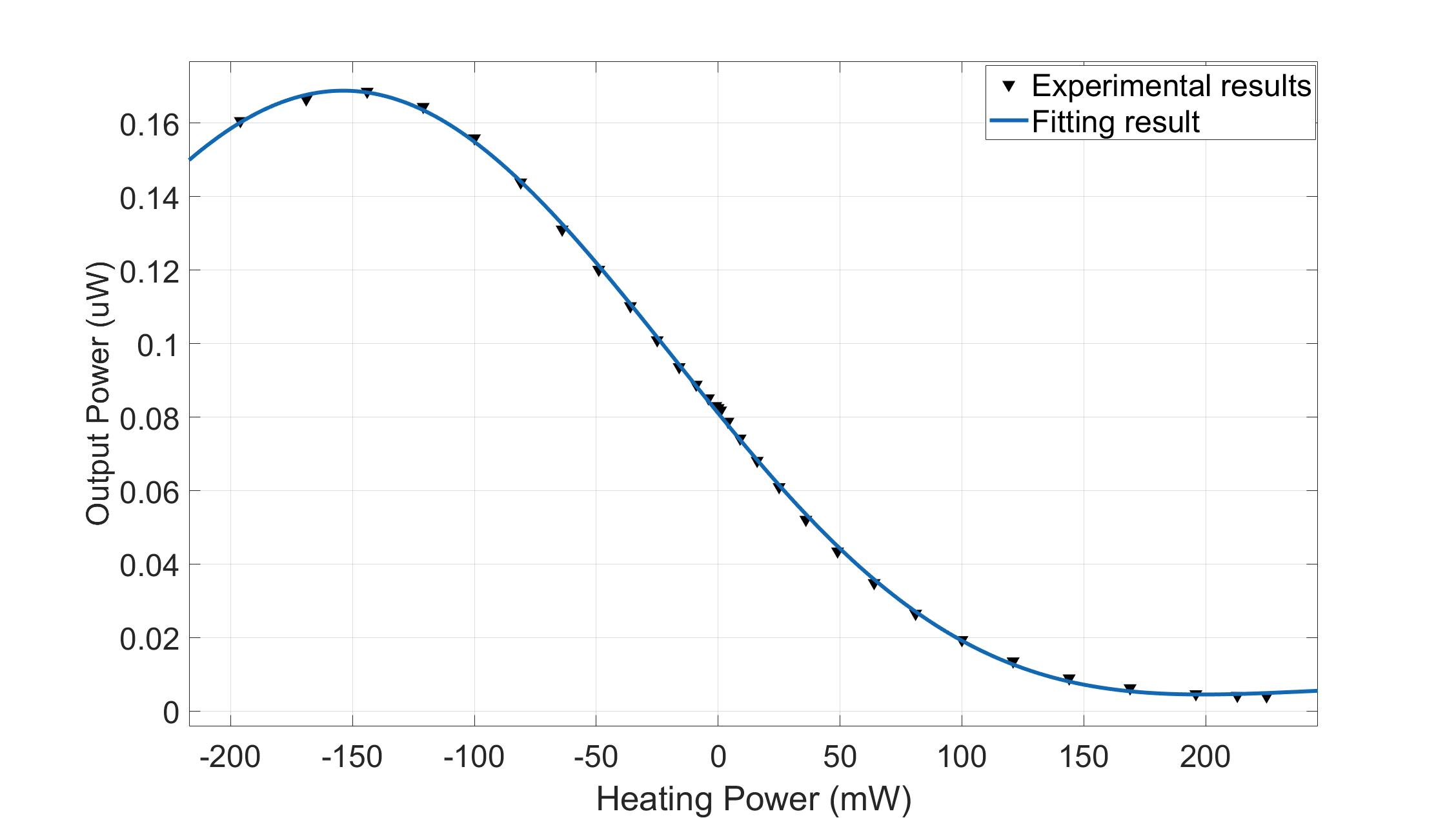}
\caption{Experimental and fitting results of the sum of waveguide output power versus the heating power. The experimental data are the concatenation of two separate runs with the heating power applied first to Waveguide 1 (the negative heating power on the plot) and then to Waveguide 3 (positive power on plot)}
\label{fig: S3}
\end{figure}

In the next step, we fit the experimental data with Eqs.~(\ref{S3}) and (\ref{S4}), with $\Delta$ being a linear function of the heating power with a baseline value $\Delta_0$.
%, as 
%\begin{equation}
%\Delta=\Delta_0+K'\times P,
%\end{equation}
%where $\Delta_0$ is a baseline value, $K'$ is the thermo-optic coefficient, and $P$ is the heating power. 
%First, the Waveguide 3 ouput is fitted with Eqs.~(\ref{S3}) and (\ref{S4}) to find $\Gamma$ . Then, the coupling strength $\Gamma$ is also adopted to fit the output from Waveguide 1 with the second equations in Eqs.~(\ref{S3}) and (\ref{S4}). 
In Fig. \ref{fig: S3}, we plot the experimental results along with the fitted curves, with the fitting parameters listed in Table. \ref{tab1}. As seen, the coupling efficiencies of the two waveguides are nearly identical. Thus in the main text, we choose their average value of 0.033, or -14.8 dB, as the coupling efficiency for both waveguides.

\begin{table}[htbp]
\centering
\caption{\bf Fitting Pamameters}
\begin{tabular}{ccc}
\hline
Parameters & Value \\
$\Gamma $(/cm) & 20.96 \\
$\Delta_0$ (/cm) & 32.95 \\
$\frac{d \Delta}{d P}$ /(mW$\cdot$ cm) & 0.21\\
$\eta_1(dB)$ & -15.23 \\
$\eta_2(dB)$ & -14.24 \\
\hline
\end{tabular}
\label{tab1}
\end{table}

\begin{figure}[ht]
\centering
\includegraphics[width=\linewidth]{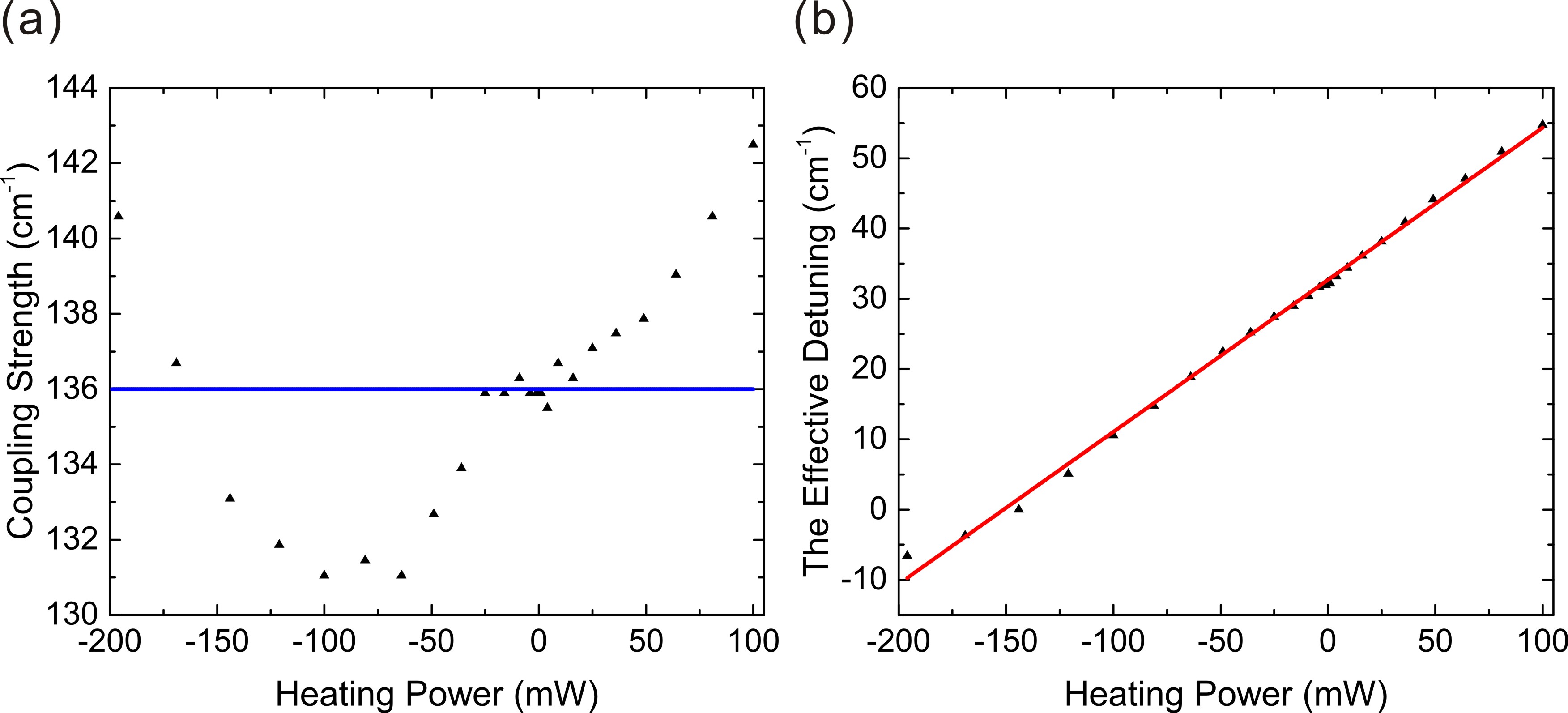}
\caption{The coupling strength and effective detuning extracted from the output power measurements. (a) The coupling strength against the heating power, where black triangles are the coupling strength and blue line is the average value of results. (b) The effective detuning against the heating power, where black triangles represent extracted results along the heating power and red line is their linear fit.}
\label{fig: S4}
\end{figure}

The coupling strength and effective detuning can also be extracted through Eqs. (\ref{S3})-(\ref{S4}) for each heating power, under the condition that the coupling efficiency ($\eta$) for both waveguides are -14.8 dB. In Fig.~\ref{fig: S4}, we present the coupling strength and the effective detuning results for varied heating power. From Fig.~\ref{fig: S4}(a), the average coupling strength is $\kappa=136 /cm$, which is used to draw the theoretical curve of eigenvalues in Fig.~3 in the main text. In Fig.~\ref{fig: S4}(b), the effective detuning reaches zero when the heating power is 144 mW. By linear fitting, the thermo-optic coefficient is 
\begin{equation}
\frac{d \Delta}{d P}=22.4*10^{-2}/(mW\cdot cm),
\end{equation}
which is close to that obtained for the single waveguide case in Eq. (\ref{S6}). This result shows that the trench is effective in stopping the thermal flow to the other waveguide.

\begin{figure}[ht]
\centering
\includegraphics[width=\linewidth]{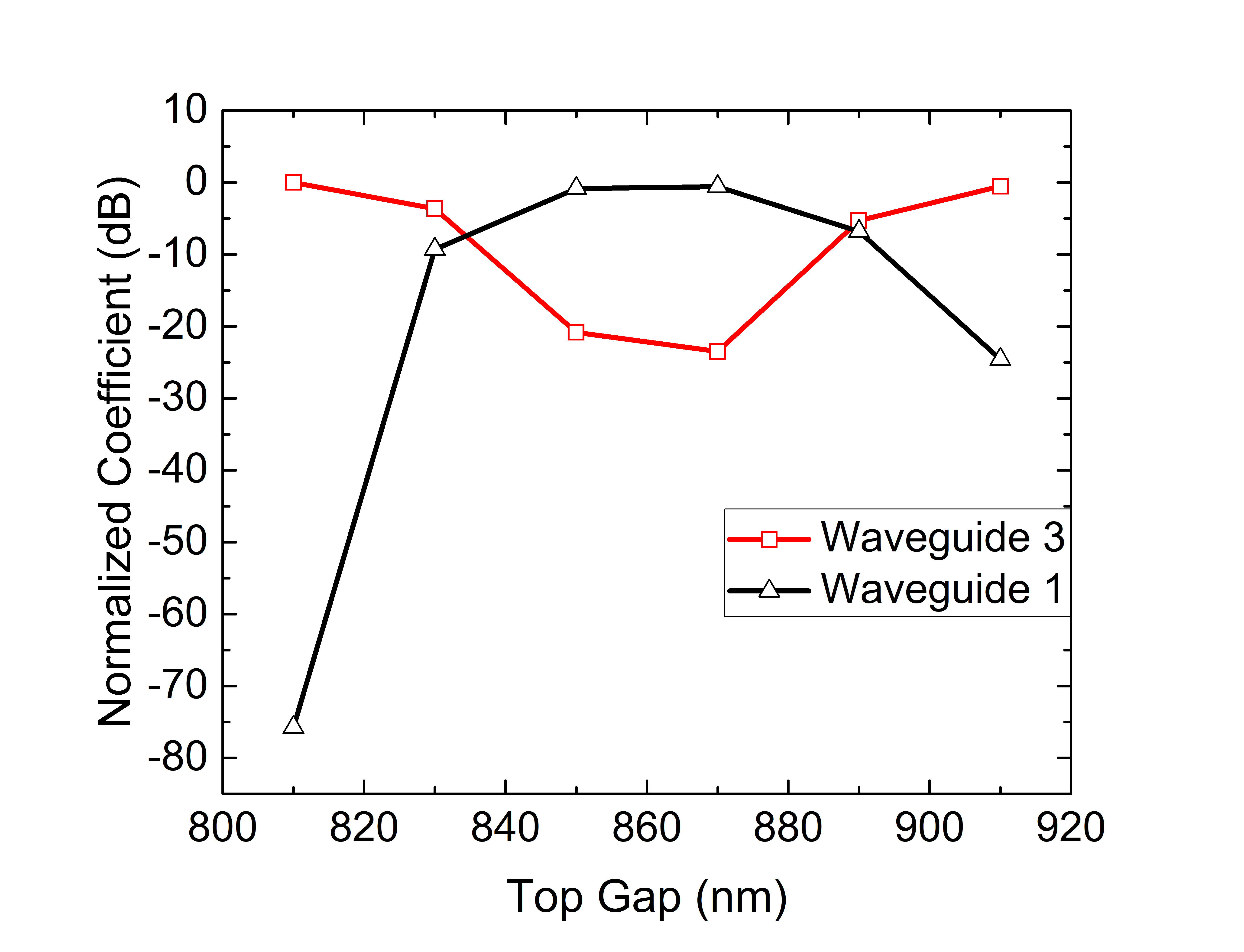}
\caption{Power splitting of a three coupled waveguides structure without the Cr strip to induce loss on Waveguide 2.}
\label{fig: S5}
\end{figure}

To show the peculiarity of the equal splitting effect observed in this anti-PT structure, we simulate a Hermitian structure of the same three coupled waveguides but without the metal strip. The dimensions of three waveguides are set the same as Waveguide 1 in the anti-PT structure. The top gap of the neighboring waveguides is gradually increased from 810 nm to 910 nm. Only Waveguide 3 is excited with an input at 1550nm. The results are shown in Fig.~\ref{fig: S5}, where the waveguide outputs are quite sensitive to the top gap. A 3-dB tolerance in power splitting would require the fabrication error to be less than 10 nm. This poses a significant challenge in fabricating photonic integrated circuits where many beamsplitters need to be nested. In contrast, in our anti-PT structure, the condition for equal power splitting is simply $ \Gamma \gg \Delta, 1/L$, which greatly increases the fabrication tolerance while offering the flexibilities of optimizing the whole circuits for other requirements. Contrasting with the results in Fig.~2 in the main text, this simulation affirms that the equal splitting observed in the anti-PT case is indeed induced by dissipation, which is quite distinct to those of conventional directional coupling ruled by the coupled mode theory \cite{Huang:94}. 

% Bibliography
\bibliography{sample}

%Manual citation list
%\begin{thebibliography}{1}
%\bibitem{Zhang:14}
%Y.~Zhang, S.~Qiao, L.~Sun, Q.~W. Shi, W.~Huang, %L.~Li, and Z.~Yang,
 % \enquote{Photoinduced active terahertz metamaterials with nanostructured
  %vanadium dioxide film deposited by sol-gel method,} Opt. Express \textbf{22},
  %11070--11078 (2014).
%\end{thebibliography}